\documentclass[sigconf]{acmart}

\AtBeginDocument{%
  \providecommand\BibTeX{{%
    \normalfont B\kern-0.5em{\scshape i\kern-0.25em b}\kern-0.8em\TeX}}}

\copyrightyear{2024}
\acmYear{2024}
\setcopyright{rightsretained}
\acmConference[CHI EA '24]{Extended Abstracts of the CHI Conference on Human Factors in Computing Systems}{May 11--16, 2024}{Honolulu, HI, USA}
\acmBooktitle{Extended Abstracts of the CHI Conference on Human Factors in Computing Systems (CHI EA '24), May 11--16, 2024, Honolulu, HI, USA}
\acmDOI{10.1145/3613905.3638195}
\acmISBN{979-8-4007-0331-7/24/05}
\begin{document}

\title{Design, Development, and Deployment of Context-Adaptive AI Systems for Enhanced End-User Adoption}

\author{Christine P Lee}
\affiliation{%
  \institution{Department of Computer Sciences, University of Wisconsin--Madison}
  \country{Madison, Wisconsin, USA}
}
\email{cplee5@cs.wisc.edu}


\begin{abstract}

My research centers on the development of context-adaptive AI systems to improve end-user adoption through the integration of technical methods. I deploy these AI systems across various interaction modalities, including user interfaces and embodied agents like robots, to expand their practical applicability. My research unfolds in three key stages: design, development, and deployment. In the design phase, user-centered approaches were used to understand user experiences with AI systems and create design tools for user participation in crafting AI explanations. In the ongoing development stage, a safety-guaranteed AI system for a robot agent was created to automatically provide adaptive solutions and explanations for unforeseen scenarios. The next steps will involve the implementation and evaluation of context-adaptive AI systems in various interaction forms. I seek to prioritize human needs in technology development, creating AI systems that tangibly benefit end-users in real-world applications and enhance interaction experiences.

\end{abstract}






\maketitle

\section{Motivation and Research Objectives}
\subsection{Motivation}


Artificial Intelligence (AI) systems are increasingly employed across diverse real-world applications, spanning home environments and various workplace sectors such as healthcare, finance, and management. Their goal 
spans from offering decision support, making decisions on users behalf, or independently generating action plans to fulfill users request. However, the introduction of AI systems has highlighted deficiencies in their interaction with users, particularly the critical issue of overlooking end-users---those directly engaging with the AI systems---during the design, development, and deployment phases. The rapid advancement of Machine Learning (ML) and AI techniques often emphasizes performance skills without a clear understanding of user and environmental needs for practical application. Recognizing that successful user use, adoption, and satisfaction with such AI systems in real-world domains require a user-centered focus, it is crucial to implement this approach from the initial design through the development stage.


Applying a user-centered approach in the technical design and development of AI systems is imperative in uncovering user problems, needs, and potential impacts on their daily lives. This understanding elicits the requirements for seamless integration into real-world applications across diverse environments, tasks, and users. Employing user-centered methods further lays the groundwork for building ``context-adaptive'' AI systems. These systems possess the ability to automatically and independently adjust their action planning and decision-making policies based on the specific environment, encompassing task requirements, user needs, preferences, and interaction patterns. Modern Machine Learning (ML) and AI techniques, including Large-Language Models (LLM) and reinforcement learning, offer a range of technical methods to support this necessary adaptivity, thereby enhancing the practical usability and personalization of AI systems. However, ensuring the reliable execution of this adaptation is also crucial, addressing concerns associated with the ``blackbox'' nature and hallucination issues. Incorporating techniques from Programming Languages (PL), including formal verification and synthesis, can be used to establish boundaries on decision-making policies and guarantee understandable and predictable outcomes.



In addition to its technical capabilities, AI systems for real-world deployment often overlook an important factor---its interaction forms. Currently, there is a notable trend in using simplistic user interfaces for AI system interactions. However, it is imperative to recognize the increasing integration of AI systems into embodied forms such as robots. This shift introduces novel paradigms, where agents engage with users through verbal and non-verbal behavioral cues while having character and social presence. Understanding the significance of interaction forms is essential in the design process, as it can greatly influence user experience, expectations, and communication preferences. To fully harness the potential of AI systems across various domains and uncover new tasks they can support, the diverse interaction forms that can be used during real-world applications must be explored.

My research focuses on building context-adaptive AI systems for end-user adoption through the integration of technical methods. I aim to deploy these context-adaptive AI systems in both disembodied and embodied interaction forms to expand the potential of AI system applications. The research involves three stages: (1) \textbf{design} through exploratory studies; (2) \textbf{development} through the integration of technical methods; and (3) \textbf{deployment} through iterative user studies for obtaining feedback and evaluating user experiences in both disembodied and embodied forms. Each stage addresses specific aspects of creating and implementing context-adaptive AI systems. 
My overarching research questions are:
\begin{itemize}
    \item [] \textbf{RQ1 (Design).} Why are context-adaptive AI systems needed for end-users?
    \item [] \textbf{RQ2 (Design).} How can end-users be involved in the design process of context-adaptive AI systems?
    \item [] \textbf{RQ3 (Development).} How can technical methods be integrated to develop context-adaptive AI systems for various interaction forms (\textit{i.e.,} simple user interfaces or embodied forms like robots)?
    \item [] \textbf{RQ4 (Deployment).} How do context-adaptive AI systems improve human-AI interactions?
\end{itemize}

\section{Related Work}
\subsection{User-Centered Focus for Practical AI and ML}
The rapid progress in ML and AI is transforming human-machine interaction and expanding the capabilities of powerful models. For example, LLMs are being deployed in applications from communicative chatbots to code generation tools to search engines \cite{liao2023ai}. ML methods such as Reinforcement Learning for Human Feedback (RLHF) \cite{casper2023open} are being used to gather expensive and time-consuming human evaluations to be done in a couple of hours \cite{dubois2023alpacafarm}. However, this advancement emphasizes prioritizing technological aspects, focusing on refining model accuracy and performance \cite{lombrozo2012explanation}. What often goes unnoticed is the critical examination of how these technological advancements will practically manifest in the real world and contribute to enhancing human experiences. The development process frequently overlooks a human-centric perspective, neglecting to scrutinize the intended purpose of the technology, its impact on users, and the implications for everyday life \cite{spektor2023designing, spektor2023charting}. This oversight can lead to a limitation in the practical utility of the developed technologies. Thus, it is crucial to center technology development around the human experience, to identify potential issues and formulate effective solutions correctly. 

Recently, Human-Computer Interaction (HCI) research has played a pivotal role in laying the foundation for understanding the use and applications of emerging technologies. The intersection of AI and HCI has evolved into a distinct research field \cite{amershi2019guidelines}, as indicated by workshops featured in premier conferences on HCI \cite{aihciworkshop2} and ML \cite{aihciworkshop}. This acknowledgment within the technical community reflects an awareness of the challenges posed by the rapid development of ML and AI, particularly in terms of human-centered considerations. Ongoing research in the ML and Explainable AI (XAI) domains is progressively embracing a human-centered perspective. Efforts are directed towards enhancing transparency in technical models \cite{liao2021human, vaughan2020human}, encompassing frameworks for documenting models and datasets \cite{crisan2022interactive, bender2018data, arnold2019factsheets}, methods for elucidating model outputs \cite{koh2017understanding, lundberg2017unified, russell2019efficient}, and strategies for effectively communicating uncertainty \cite{bhatt2021uncertainty, dhami2022communicating}. As AI and ML models continue to integrate into daily life, active engagement of HCI research in the design and development of these technical processes becomes paramount. A human-centered lens is indispensable to ensure that these systems are not only technologically advanced but also practical, reliable, and genuinely beneficial to human users in diverse contexts. This coherent approach is vital for bridging the gap between technological innovation and its meaningful integration into human experiences.


\subsection{User-Centered AI systems in HCI}

To ensure the development of technology is user-centric, a comprehensive understanding of end-users and their deployment environment is essential. A significant body of research has focused on using design methods and tools to engage diverse stakeholders throughout the AI system design process~\cite{lee2019webuildai, zhang2023deliberating, lee2021participatory}. \citet{lee2019webuildai} introduced a framework that facilitates the participation of various stakeholders in building algorithmic policies that align with their interests. In relation, \citet{zhang2023stakeholder} demonstrated the effectiveness of co-design sessions with users, incorporating data probes to capture the perspectives and needs of workers. 
Moreover, efforts have been directed toward enhancing the explainability of AI systems by understanding user and contextual needs. \citet{liao2021human} delved into the distinct needs of different user types, such as developers, business owners, and decision-makers. Similarly, consideration of situational factors of past interactions with others in the design of human-centered AI systems has been emphasized \cite{ehsan2021expanding}. Understanding these human-centered factors is crucial for enhancing user trust and adoption, and ensuring successful integration of AI systems \cite{liao2022designing, lee2021included, ehsan2019automated}.

Beyond the technical aspects, researchers have drawn insights from related fields, such as social sciences, human psychology, and communication sciences, to comprehend how humans respond to interactions with AI systems. Prior work has explored the psychological impact of data-driven management algorithms in the workplace \cite{lee2015working, zhang2022algorithmic}. \citet{wang2019designing} extended this interdisciplinary approach by connecting human cognitive theory to the design of AI systems, aiming for improved communication aligned with the typical human thought process. Further, \citet{miller2019explanation} brought insights from social sciences into the design of AI explanations, considering of fundamental attributes of human intellect for the development of human-centered technology development.

\section{Contributions to Date}

\subsection{Exploring Design Needs of Context-Adaptive AI systems for Improved Human-AI Interaction \cite{lee2024workplaceAI}}\label{sec:s1design}

This study aimed to understand real-world user experiences of interacting with AI systems in their daily activities. 
We specifically collaborated with professionals in finance, healthcare, and management who regularly utilized AI systems in their work settings to comprehend their day-to-day experiences with these systems. The AI systems in these workplaces served as decision-support tools or autonomously made and distributed decisions to the employees.
The research revealed a significant gap between the offerings of AI systems and the needs and preferences of the users. 
Despite the similar functionality of decision-support, the AI systems played different roles in the workplace. Specifically, in the healthcare domain, the AI system played an assistive role that involved follow-up tasks wherein workers used the AI output for other related tasks. In the finance domain, the AI system played a decision-making role and involved similar follow-up tasks. In the management domain, the AI system played a decision-making role and did not involve any related follow-up tasks for the workers. 
The distinct roles of the AI system in workplaces provided different levels of worker autonomy. The level of autonomy combined with the unique workplace characteristics and worker values, led to different sets of challenges and information needs for the AI system. 
For workplaces where the AI system had a more assistive role and provided higher worker autonomy, information needs focused on expanding AI's informative capabilities to support and improve worker task performance. In workplaces where the AI system predominately held a decision-making role rather than supporting worker autonomy, worker information needs focused on enhancing their autonomy and opportunities for exercising worker expertise. Based on our, findings, we present design implications for context-adaptive AI systems to consider their role in the deployed environment, situational factors, and user values for successful integration.



\subsection{Incorporating End-Users in the Design of AI Systems \cite{lee2024dexai}}


Expanding on the insights presented in Section \ref{sec:s1design}, we developed a design tool to empower end-users to actively contribute to the design process of AI systems and conducted co-design sessions for evaluation of the tool. This user-centered approach was informed by my prior work, which involved similar co-design sessions with end-users to enhance their initial experiences when interacting with a robotic agent \cite{lee2022unboxing, lee2023demonstrating}. The goal of developing a user-centered design tool was to incorporate end-users in the design process of AI systems, enabling them to articulate their needs and actively shape the features, communication methods, and interaction procedures. The result of this effort was the ``DeX-AI,'' a set of design cards rooted in communication and organization theory as presented by \citet{mohr1990communication}. The DeX-AI facilitates constructive communication of users' explanation needs for AI systems through four distinct communication channels: content, modality, frequency, and direction. Each category includes design elements inspired by XAI and design literature and a blank card for users to generate their own designs. Functioning as a design tool for end-users to be used within traditional HCI research methods, the DeX-AI enabled users to prototype explanations tailored to their context-specific requirements. The end results offered tangible examples to inform the communication and explanation methods of the AI system.

We conducted co-design sessions where workers used the DeX-AI to prototype ideal AI explanations. During the follow-up interviews, participants reported that the design cards helped clarify their abstract needs, as the constructive nature of the design cards required participants to systematically analyze their interaction with the AI system. The design cards proved valuable in expressing their requirements and preferences related to task distribution between the AI system and the worker, the desired design of AI explanations with considerations for others' interactions, and support for individual task planning. Notably, participants expressed a desire for the AI system to adapt its decision-making or communication policies based on user feedback and data collected during interactions, for continuous alignment with their evolving needs.


\subsection{Development of Context-Adaptive AI Systems for Robotic Agents \cite{lee2024rex, kim2024understanding}}

Building on empirical findings derived from observation and design studies, my focus shifted toward the development of context-adaptive AI systems. I selected a commonplace, everyday use case where technology needed to constantly adapt based on context, leading to the selection of robot agents assisting with various household chores in a home setting. The goal of the AI system was to autonomously generate alternative solutions to situations where the robot agent encountered unforeseen challenges in fulfilling user requests, providing comprehensive explanations for both the conflict and the generated solution. Furthermore, we aimed to enhance the resilience and safety of context-adaptive AI systems, addressing the challenges posed by their opaque nature in output generation during user interactions. Thus, we designed an AI system rooted in PL methods, capable of automatically detecting conflicts, generating relevant alternative actions as solutions, and providing users with explanations detailing the conflict and solution. Formal PL methods were employed to guarantee the safety of the generated alternative solutions and actions, including program synthesis and verification. Program synthesis entails automatically generating an executable program based on partial information and high-level specifications, while program verification employs mathematical modeling to assess the program against predefined safety properties, ensuring alignment with specified correctness criteria.

Findings from both quantitative and qualitative aspects of the user study indicated that automated repair and explanations were valuable. Explanations providing insights into both the conflict and the alternative solution proved most effective in enhancing users' understanding of the robot's actions, ultimately increasing satisfaction and willingness to further engage with the robot. Additionally, we identified specific risk factors requiring additional consideration during automated repair by the robot. Our findings outline users' expectations and needs regarding appropriate robot actions and explanations in scenarios involving these risk factors. Finally, we provide design implications for the implementation of automated repair and explanations by AI systems for autonomous agents. Additionally, we discuss future work on exploring how PL-based AI systems can be seamlessly integrated with other AI techniques, such as LLMs.

\section{Expected Next Steps}

\subsection{Integrating Technical Methods for Development of Context-Adaptive AI Systems}
In my next work, I plan to integrate additional technical methods to elevate the capability of the context-adaptive AI systems to be applied to a wider range of applications and interaction forms. Specifically, I will focus on enhancing the capabilities of the context-adaptive AI system with powerful and robust technical methods. I will integrate methods from ML for adaptability, PL for safety, and eXplainable AI (XAI) for interpretability. Using these methods, end-users will be able to adapt the AI system across three phases: the Synthesis and Verification phase for policy-level changes in system design, and the Repair phase for output-level changes during system runtime. During the \textit{Synthesis} phase, the AI system will use Reinforcement Learning from Human Feedback (RLHF) \cite{griffith2013policy} and automated planning \cite{ghallab2004automated} to adapt its decision-making policy to user needs. RLHF involves three stages: gathering user feedback on system outputs, reward modeling through supervised learning, and optimizing the AI policy for improved performance. User feedback will inform AI system design changes, such as specifying task distributions, decision-making logic, and support needs. The AI system will integrate this feedback and past interaction data to update its model. To ensure the safety of policy-level design changes in the AI system, the  \textit{Verification} phase will employ PL verification methods \cite{baier2008principles} to assess the model's alignment with specified correctness criteria. Models not meeting the verification conditions will be ineligible for inclusion in the AI system. Users will also be involved in validating and defining correctness properties. Upon the completion of adapting the AI system design, it will engage users through XAI techniques, providing insights into model behavior and alternate outcomes until a new policy-level design request emerges. During runtime, the \textit{Repair} phase will address changes related to the AI system’s outcomes. First, automatic repair will use PL repair methods \cite{jobstmann2005program} to rectify user errors and minor technical issues, such as security vulnerabilities and resource allocation. PL repair methods involve correcting faults with partial specification, such as when a user underspecifies a low-stakes term in their request. Conversely, interactive repair \cite{holtz2018interactive} will use RLHF for users to correct the system's output errors.

\subsection{Evaluating End-User Experiences in the Deployment of Context-Adaptive AI Systems}

Concurrently with model development, iterative user studies will assess user experiences with the context-adaptive AI system. These human-centered studies will offer insights into benefits, deficiencies, and user needs for AI system adoption and how to integrate contextual requirements. Throughout each technical development phase, multiple user studies will examine user engagement with specific aspects of the system. For instance, during the \textit{Synthesis} and \textit{Verification} phase, studies will explore when and how users most prefer to provide feedback and its impact on their perceptions of the AI system. User studies will also involve end-users to participate in the design of the AI system, through brainstorming sessions and vision creation using design tools, probes, and storyboarding methods to envision the holistic interaction process with the AI system. Both quantitative and qualitative measures will be gathered to support the findings of each user study.


Additionally, the context-adaptive AI system will be deployed in both embodied and disembodied interaction forms. First, the disembodied interaction form will undergo user studies in real-world settings, specifically in knowledge work domains where information communication between the AI system and the user is the central task. Conversely, AI systems with embodied interaction forms will be deployed for tasks necessitating social and physical engagement, such as robots assisting with household chores in a home setting. These studies for distinct interaction forms will underscore varied needs, suitable deployment environments, and the required design enhancements for the future development of context-adaptive AI systems tailored to each interaction form.


\section{Anticipated Contributions}


The rapid deployment of AI systems highlights a gap between user needs and system capabilities, necessitating research to design and develop AI systems with a user-centric focus. My work integrates human-computer interaction with technical skills, prioritizing a ``humans first'' approach. The anticipated outcomes of the proposed research projects include: (1) advancements in the integration of technical methods from Machine Learning, Programming Language, and eXplainable AI to construct context-adaptive AI systems; (2) the development of practical AI tools and systems applicable across diverse domains, capable of integration with existing tools; and (3) the formulation of guidelines for incorporating end-user needs and context in the design and development processes of AI systems. I believe this research focus is critical in unlocking the full potential of AI systems for various applications, especially when AI systems are becoming an indispensable asset for everyday domains.

\section{Dissertation Status and Long-term Goals}
I am a second-year Ph.D. student in the Computer Sciences Department at the University of Wisconsin-Madison, under the guidance of Dr. Bilge Mutlu. I finished the required coursework in Computer Science, passing my Ph.D. qualification exam in the Summer of 2023. I aim to present my dissertation by Spring 2026. Post-graduation, my goal is to engage in HCI research as a full professor. My research, rooted in technical methods such as Machine Learning and Programming Languages, focuses on enhancing human-AI interaction in both embodied and disembodied forms. This aims to contribute to the development of practical and trustworthy AI systems for real-world applications. Participating in the CHI'24 Doctoral Consortium will be a valuable experience to gain insights and mentorship from fellow researchers, as I have not previously attended a doctoral consortium at any other SIGCHI event.

\begin{acks}
I am grateful to my advisor Professor Bilge Multu and collaborator Professor Min Kyung Lee at the University of Texas--Austin for their continuous mentorship and support. Additionally, I thank the People \& Robots Lab at the University of Wisconsin--Madison and the participants in user studies for generously sharing their time and experiences.
\end{acks}


\bibliographystyle{ACM-Reference-Format}
\bibliography{bibliography}

\end{document}